*In silico* evolution of signaling networks using rule-based models: bistable response dynamics*

Song Feng[1], Orkun S. Soyer[2]

1. Center for Nonlinear Studies, Theoretical Division, Los Alamos National Laboratory, United States
2. School of Life Sciences, University of Warwick, United Kingdom. Email: O.Soyer@warwick.ac.uk

**Abstract**
One of the ultimate goals in biology is to understand the design principles of biological systems. Such principles, if they exist, can help us better understand complex, natural biological systems and guide the engineering of *de novo* ones. Towards deciphering design principles, *in silico* evolution of biological systems with proper abstraction is a promising approach. Here, we demonstrate the application of *in silico* evolution combined with rule-based modelling for exploring design principles of cellular signaling networks. This application is based on a computational platform, called BioJazz, which allows *in silico* evolution of signaling networks with unbounded complexity. We provide a detailed introduction to BioJazz architecture and implementation and describe how it can be used to evolve and/or design signaling networks with defined dynamics. For the latter, we evolve signaling networks with switch-like response dynamics and demonstrate how BioJazz can result in new biological insights on network structures that can endow bistable response dynamics. This example also demonstrated both the power of BioJazz in evolving and designing signaling networks and its limitations at the current stage of development.

**Introduction**
Biological systems are complex. This complexity derives from combinatorial interactions between the components of biological systems across and within multiple scales *(1-5)*. These complex interactions cannot be understood without systematically investigating the dynamics of corresponding systems *(6-8)*. Systems biology emerged as cutting edge area to decipher and understand the complexity of biological systems by integrating mathematical models with experimental data collection*(9-11)*. In parallel, synthetic biologists focus on building *de novo* biological systems, in order to test gained knowledge within well-defined, constructed systems *(12-15)*. At the core of these efforts is a desire to decipher key design principles, which can be applied broadly to understand diverse sets of biological systems. Such design principles could help characterize the mapping between the structures of cellular networks and their dynamics. For instance, as cellular information processing systems, different structures of signaling networks give rise to diverse input-output functions that can generate different response dynamics from specific input signals *(7)*. With discovered design principles, we will be able to better understand this diversity in the structure-function mapping and even predict which cellular features will lead to what type of mapping.

From an evolutionary perspective, the complexity of biological systems can be understood as emerging from combinations of adaptive and non-adaptive processes *(16, 17)*, with the former occuring under simple or fluctuating selective pressures *(18, 19)*. Evolution can be seen as moving genotypes within a structure-dynamics mapping, where genotypes encoding for different network structures result in different dynamical capabilities, which would contribute to organismal fitness (Figure 1). From this evolutionary stance, one can imagine two approaches that can be useful towards understanding design principles of biological systems; we can systematically compare components and interactions of cellular networks in different species such that we can sketch the evolutionary landscape and look for emerging trends in this landscape that can serve as design principles; alternatively, we can start from different networks and forwardly evolve them to map the evolutionary landscape. Both approaches can inform design principles of naturally occurring systems and also help to discover potential designs that we have not been explored yet *(20-23)*.

Applying computational modelling and *in silico* evolution of cellular networks can be a promising route towards exploring and discovering their design principles, however, there are significant challenges in modelling and evolving such networks and in particular signaling networks. Signaling networks are composed of a range of different types of proteins such as receptors, adaptor proteins, kinases, transcription factors and secondary messengers like calcium and nitric oxide. These proteins usually adopt multiple conformational states that allow them to carry out different binding interactions or enzymatic reactions and consequently affect other, "downstream" proteins. Through combined effect of such interactions signals are detected, encoded, integrated and transformed into changes in the activity of transcription factors, through which the cell responds in form of altering gene expression. Alternatively, signals are conveyed as modification of molecular machines controlling cellular physiology, such as movement or generation of neural action potential *(7, 24)*. In recent years, the textbook view of signaling networks being linear cascading pathways has been replaced by the emerging viewpoint of combinatorial networks formed with interconnecting proteins that have multiple binding domains or phosphorylation sites. Many signaling proteins and molecules share downstream or upstream interaction partners, which brings about "cross-talks" between different signaling "pathways" *(25)*. Given this complexity, the conventional modeling method that uses differential equations does not provide an ideal solution for simulating *in silico* the evolution of signaling networks. The primary challenge in modeling expanding (or growing systems such as evolving signaling networks) is the "combinatorial explosion" in numbers of both reactions and species *(26)*. This makes changing and restructuring signaling networks difficult to implement within the modeling approaches that are solely based on differential equations *(26)*.

In order to perform *in silico* evolution of signaling networks, we implemented a computational platform called BioJazz, which describes the signaling networks with rule-based models to overcome these difficulties. The benefits of rule-based approach in evolving signaling networks are significant. Firstly, it uses rules to describe interactions between proteins, allowing condensing the information required to encode reactions. Compared to using just (and directly) the differential equations to describe the dynamics of each signaling entity, the rule-based approach allows the use a low number of rules to capture a large network composed of re-occuring reaction types. Secondly, it is more convenient to manipulate or change the rules, such as adding, deleting or transforming proteins/domains as well as interactions. As a result, the description of signaling networks within a rule-based model allows these readily to evolve and be restructured *(20)*. The rules can then be transformed into a differential equation based model, allowing simulation and analysis of dynamics. In the following, we provide an overview of the structure of BioJazz and explain in detail its use for *in silico* evolution experiments by selecting for signaling networks with defined dynamics. These simulations result in the evolution of signaling networks with bistable response dynamics, illustrating how BioJazz can be used to decipher key structural determinants of specific network dynamics.

**Software**

BioJazz is implemented as a platform to perform *in silico* evolution of rule-based models towards studying the design principles of the signaling networks. In this approach, we forward-simulate the evolutionary process of signaling networks (Figure 2A) under different environments. The main features of BioJazz are: (1) it evolves both network topology (i.e. interactions) and connection weights (i.e. parameters); (2) it designs a network "*de novo*" (i.e. starting from randomly generated starting networks), or from user-specific "seed" networks; (3) it can be parallelized to speed up the evolutionary simulation; and (4) it is highly configurable.

This software utilizes a well-defined yet flexible encoding scheme to characterize the structure of signaling networks at three different abstraction levels: binary strings, reaction rules, and differential equations (Figure 2A) *(20)*. All three forms of abstraction are one to one mapping and all information of the network structure, including the interactions and their parameters, are preserved when the network is transformed from one to the other. It can be noticed that this three-tier encoding is akin to that found in Nature, where we move from an abstract binary encoding (representing DNA-based encoding), to a reaction-based encoding (representing biochemical reactions), to an encoding allowing dynamical simulations (representing network function). By encoding the network in this "natural" fashion, BioJazz structure allows us to represent a variety of mutational operators. Thus, using the binary encoding level, we can readily introduce point mutations, gene duplications/deletions, domain duplications/deletions, and domain shuffling *(20)*. The details of the encoding and decoding schemes between reaction rules and binary strings are introduced in reference *(20)*, while compiling from reaction rules to differential equations as well as simulation of compiled differential equations are explained in Allosteric Network Compiler (ANC) *(27)* and Facile *(28)*.

The overall evolutionary algorithm is illustrated in Figure 2B. First, we configure the evolutionary processes by defining all the parameters used for the simulation; then instantiate the starting networks by either a randomly generated one or specifically synthesized one, which is determined in the configuration file. After initialization, the evolutionary process starts by mutating the "seed" network, then evaluates the new networks and selects for the next generation. This evolutionary process iterates the "mutation-compiling-simulation-selection" cycle until certain criteria are met, such as reaching a maximum number of generations or threshold fitness score. (Figure 2B). These criteria are defined in the configuration file.

To install BioJazz, one can simply download the source code (http://oss-lab.github.io/biojazz/) or use `git` to clone the source code (`git clone https://github.com/OSS-Lab/biojazz.git`) (*see* **Note 1**), then put the source code folder into the designate directory, for instance `~/workspace/biojazz`. In addition, BioJazz requires both ANC and Facile to compile the reaction rules (as well as Matlab to solve the compiled differential equations). All three software are written in Perl with command-line interface only for consistency. It is recommended to install the software either on Linux or Mac OS X in order to ease the install and configuration efforts. Detail installation guide are included with BioJazz, as well as separately for ANC and Facile manuals (http://anc.sourceforge.net/wiki/index.php/Allosteric_Network_Compiler)                ,

(http://facile.sourceforge.net/wiki/index.php/Main_Page) (*see* **Note 1**). Additional notes for configuring the shell environment are provided as well (*see* **Note 2**). If the user wants to run BioJazz in a computer cluster, it is recommended to use clusters with multiple nodes and at least 2GB memory allocated to each node.

**Methods**

Initiating and running an evolutionary simulation in BioJazz requires two user-provided files; a configuration file describing all user-settable parameters for evolutionary simulations, and a `.pm` file describing the fitness function to be used for selection criteria on network dynamics. In order to run BioJazz, the user should have both files ready. In the following subsection, we will explain them in detail and take the ultrasensitive fitness function *(20, 21)* as an example for demonstrating evolutionary simulations in BioJazz.

*Create the workspace*

Depending on your specific application, BioJazz will require some customized configuration and scoring functions. Also, during a single simulation run, BioJazz will generate large number of output files. For this reason, the user must create a properly configured workspace, which will contain the appropriate input and output files. To facilitate this, BioJazz can create the workspace for you and populate it with the required directories and with template files to get you started. To do this, run the following command: `biojazz --command='create_workspace("bjazz")'`. This will create the directory `bjazz` and various sub-directories including `config` and `custom`. The configuration files go in the `config` directory, while the customized scoring functions are in the `custom` directory. At this point, the user should get familiar with some of the template files that are provided, and try to run BioJazz.

*Customize the scoring function and starting networks*

Before configuring and starting an evolutionary simulation, the user needs to specify which scoring function to use. Normally, this specific function is determined by the objective of the evolutionary simulations. Whether to study evolution of signaling networks with adaptive response dynamics or to design signaling networks with oscillatory dynamics, the user needs to implement and use appropriate scoring functions. In some cases, implementing an appropriate scoring function requires many test simulations. Since signaling networks can evolve with unlimited complexity in BioJazz *(20)*, different parameters in the scoring function might result in the evolution of completely different results and designs. By default, the current version of BioJazz provides two successful fitness functions: ultrasensitive (i.e. switch-like) response function and adaptive response function *(20, 21)*. (*see* **Note 3**)

At the end of each scoring function file, the user can specify a designed network by defining each section of the binary string. One can copy the current designed "seed" networks in `ultrasensitive.pm` or `adaptive.pm` files *(20, 21)* and modify them into any starting networks. Alternatively, if the user does not define a "seed" network, BioJazz can generate a randomly generated network by compiling from a random binary string. The choice of starting networks and length of the random binary string can be defined in the configuration file. An example seed network is shown in Figure 3A.

*Setting up configuration file*

The configuration files are located in the "config" folder. Each configuration file corresponds to one particular scoring function (see the previous subsection and **Note 3**). One can use some of these sample configuration files to test and understand how each of the user-settable parameters will affect the evolutionary simulations and consequently the results. The typical configuration file is shown in **Listing 1**. In the following paragraphs, we will explain some important parts of the configuration file and their functions in evolutionary simulations *(20)*.

**Listing 1.** The configuration file with all parameters required to run evolutionary simulations. Any line with # at initial is comment and will not be read by BioJazz.

```
#---------------------------------------
# CPU AND CLUSTER SETTINGS
#---------------------------------------
cluster_type = LOCAL
cluster_size = 1
nice = 15
vmem = 200000000
#---------------------------------------
# WORKSPACE AND CUSTOM SCORING MODULES
#---------------------------------------
scoring_class = Ultrasensitive
work_dir = ultrasensitive
local_dir = ultrasensitive/localdir
initial_genome = load test/custom/Ultrasensitive.obj
#---------------------------------------
# GENOME PARAMS
#---------------------------------------
# Genome class
radius = 3
kf_max = 1e3
kf_min = 1e-3
kb_max = 1e3
kb_min = 1e-3
kp_max = 1e3
kp_min = 1e-3
# Gene class
regulated_concentration_width = 10
gene_unused_width = 4
regulated_concentration_max = 1e3
regulated_concentration_min = 1e-3
# Domain class
RT_transition_rate_width = 10
TR_transition_rate_width = 10
RT_phi_width = 10
domain_unused_width = 4
RT_transition_rate_max = 1e2
RT_transition_rate_min = 1e-2
```

```
TR_transition_rate_max = 1e2
TR_transition_rate_min = 1e-2
RT_phi_max = 1.0
RT_phi_min = 0.0
# ProtoDomain class
binding_profile_width = 10
kf_profile_width = 20
kb_profile_width = 20
kp_profile_width = 10
steric_factor_profile_width = 20
Keq_profile_width = 10
protodomain_unused_width = 4
Keq_ratio_max = 1e2
Keq_ratio_min = 1e-2
#---------------------------------------
# EVOLUTION PARAMS
#---------------------------------------
num_generations = 10000
target_score = 0.8
first_generation = 0
continue_sim = 0
continue_init = 0
remove_old_files = 1
score_initial_generation = 1
rescore_elite = 0
report_on_fly = 1
report_selection = 0
# selection method: kimura selection
selection_method = kimura_selection
effective_population_size = 1e8
amplifier_alpha = 1e3
max_mutate_attempts = 100000
# selection method: population-based selection
#selection_method = population_based_selection
#fossil_epoch = 10
#inum_genomes = 50
#evolve_population = 1000
#mutation_rate = 0.05
# mutation settings
```

```
mutation_rate_params = 0.0
mutation_rate_global = 0.01
gene_duplication_rate = 0.005
gene_deletion_rate = 0.005
domain_duplication_rate = 0.005
domain_deletion_rate = 0.005
recombination_rate = 0.01
hgt_rate = 0.01
#---------------------------------------
# ANALYSIS PARAMS (POST-EVOLUTION)
#---------------------------------------
restore_genome = 0
analysis_dir = analysis
#---------------------------------------
# ANC PARAMS
#---------------------------------------
max_external_iterations = -1
max_internal_iterations = -1
max_complex_size = 3
max_species = 512
max_csite_bound_to_msite_number = 1
default_max_count = 2
default_steric_factor = 1000
export_graphviz = network,collapse_states,collapse_complexes
# FACILE/MATLAB SETTINGS
solver = ode23s
sampling_interval = 1.0
SS_timescale = 500.0
# MATLAB odeset params
InitialStep = 1e-8
AbsTol = 1e-9
RelTol = 1e-3
MaxStep = 500.0
#---------------------------------------
# SCORING PARAMS
#---------------------------------------
plot_input = 1
plot_output = 1
plot_species = 0
```

```
plot_phase = 1
plot_min = -1
round_values_flag = 0
steady_state_threshold = 1000
steady_state_score_threshold = 0.5
delta_threshold = 0.01
amplitude_threshold = 0.01
ultrasensitivity_threshold = 5
complexity_threshold = 250
expression_threshold = 500
w_n = 0.0
w_c = 0.0
w_e = 0.0
w_s = 1.0
w_a = 1.0
w_u = 1.0
w_u1 = 1.0
w_u3 = 1.0
LG_range = 10
LG_delay = ~
LG_strength = 4.0
LG_ramp_time = 3000
LG_steps = 3
LG_timeout = 20000
stimulus = ss_ramp_equation
hill_n = 40
hill_k = 5
TG_init = 1000
cell_volume = 1e-18
lg_binding_profile = 0100111010
tg_binding_profile = 0111000110
# SPREADSHEET EXPORT/ANALYSIS
genome_attribute_names = score, ultrasensitivity_score, expression_score,
amplitude_score, complexity_score, steady_state_score, complexity,
num_anc_species, num_rules, num_genes, num_pruned_genes, num_domains,
num_protodomains, num_allosteric_domains, num_allosteric_protodomains,
num_binding_protodomains, num_phosphorylation_protodomains,
num_catalytic_protodomains, num_kinase_protodomains,
num_phosphatase_protodomains, num_adjacent_kinases,
num_adjacent_phosphatases, num_receptive_protodomains, tg_K1, tg_K2,
tg_K1_concentration, tg_K2_concentration
```

CPU AND CLUSTER SETTING: This block defines the number of nodes and amount of memory that will be allocated to BioJazz. This setting is vital for simulations with `population_based_selection` method *(20)*, as this method utilizes a parallelization algorithm that allows scoring a group of "genomes" simultaneously. The `cluster_size` determines how many individuals in each generation can be scored in parallel. With `kimura_selection` method (see EVOLUTION PARAMS), the parallelized scoring is not utilized and therefor `cluster_size` is normally set as 1. The `vmem` parameter determines how much virtual memory will be allocated to Matlab for solving the differential equations. This value should be not less than 200000000 (i.e. 200MB).

WORKSPACE AND CUSTOM SCORING MODULES: This block is used to specify the name of the fitness function, directory of simulation results and locations of files containing starting networks. The `scoring_class` defines the name of the file describing the fitness function, in this case *Ultrasensitive*. When BioJazz starts scoring individual networks, it looks for a Perl module file named with `Ultrasensitive.pm` in the folder `custom` to execute the fitness function. The `work_dir` defines the directory where BioJazz will put the simulation results, which can be any valid Linux/Unix folder name. The `local_dir` is the directory where all temporary files generated during a simulation are placed. The `initial_genome` specifies the defined starting network or can be set as `random`, if the user intends to start simulations from a randomly generated network. The randomly generated network is compiled from a random binary string with length 5000 (this value is currently defined in source code of BioJazz).

GENOME PARAMS: This block defines all required parameters governing the conversion between the genome-like binary encoding of a network and the corresponding rule-based model. The former has a hierarchical structure as illustrated in Figure 3B. There are four subblocks defining all levels of the binary string necessary to encode a network; "genome", "gene", "domain", and "protodomain". The "genome class" contains the parameters relating to the overall behavior of the rule based model, including reactivity and reaction parameters *(20)*. In order, these parameters are: `radius`, which defines the threshold whether two "reactive sites" can interact with each other *(20)* (*see* **Note 4**), and the following 6 parameters (i.e. `kf_max`, `kf_min`, `kb_max`, `kb_min`, `kp_max`, `kp_min`) define the range (i.e. maximum and minimum) of reaction rate constants. The "gene class" contains parameters related to individual proteins. For instance `regulated_concentration_width` defines the width of binary string that encodes the concentration of a corresponding protein, and both `regulated_concentration_max` and `regulated_concentration_min` together determines the range of concentration parameters. The appropriate ranges of reaction rate constants and concentration parameters are explained in detail in previous applications of BioJazz *(20)* and *(21)*. The "domain class" and "protodomain class" section contain parameters related to allosteric flags, types of reactive site as well as encoding profiles of reaction rate constants (see detailed explanations in BioJazz manual and previous publications *(20)* and *(21)*). All of the parameters in these subblocks can be set by the user. The default parameters of length for calculating concentrations and reaction rate constants, such as `regulated_concentration_width`, `RT_transition_rate_width`, `binding_profile_width`, `kf_profile_width`, determine how dense of the reaction parameters that are allowed to mutate in defined ranges. The user can also tune these settings to explore a finer parameter space.

EVOLUTION PARAMS: This block defines the parameters required for controlling the evolutionary algorithm. In particular, it sets the initiation, termination, mutations, and selections. In order to make evolutionary simulations, several parameters in this section should be tuned to suit the needs of the objectives. The `num_generations` sets the maximum number of generations the simulation are allowed to reach, similarly the `target_score` sets the cut-off on the fitness score; the evoltuioanry simulation will terminate once it reaches any one of such conditions. The `selection_method` are used to defined selection methods of the evolutionary algorithm among the two available approaches; `kimura_selection` and `population_based_selection` *(20)*. The associated parameters of each selection method (i.e. `effective_population_size`, `amplifier_alpha` and `evolve_population`, `mutationa_rate`) are defined in the selection equations *(20)*. At the end of this block are mutation rates for different types of mutations. The `mutation_rate_params` defines rate of point mutations only on reaction rate constants and concentrations. This parameter is particularly important if one wants to freeze the structure of networks but optimize the parameters. The `mutation_rate_global` defines the rate of point mutations at a global scale (i.e. the whole binary string). The `gene_duplication_rate`, `gene_deletion_rate`, `domain_duplication_rate`, `domain_deletion_rate` and `recombination_rate` are used to define the mutation rate for gene/domain duplication or deletion as well as domain recombinations.

ANC, FACILE, AND MATLAB SETTINGS: This block defines the parameters for compiling the ANC rules into differential equations. Most of these parameters can be referred in ANC documentations (http://anc.sourceforge.net/wiki/index.php/Main_Page) and Facile documentation (http://facile.sourceforge.net/wiki/index.php/Main_Page).

SCORING PARAMS: This block contains parameters determining the behavior of scoring functions. These parameters are set in a unique way for each of the different scoring functions that can be generated by the user as a Perl file. To better understand these, the user can use the example scoring function files, available in the `custom` folder. Reading these files requires basic familiarity of Perl programming language. In the sample ultrasensitive.pm, we have defined several parameters relating to the implementation, evaluation and visualization of network fitness based on response dynamics. For example, `plot_input`, `plot output`, and `plot species` defines whether the simulation will produce the time-course plots of input signal, output response as well as each species in each generation; `steady_state_threshold`, `amplitude_threshold`, `ultrasensitivity_threshold`, and `complexity_threshold` are used to define the thresholds for different scores regarding steady states behavior, response amplitude, ultrasensitive level, and complexity level of the network; `w_s`, `w_a`, `w_u`, and `w_c` are used to define the weights of these four scores contributing to the final fitness score. Also the `stimulus` and `TG_init` are used to define the type of input signals and maximum of response level (i.e. the total concentration of output protein). All these parameters can be changed, or new parameters can be introduced, by the user and according to the selection function that they would like to implement.

*Running the simulations*
After installing all required Perl modules (*see* **Note 1** and **2**), one can run BioJazz to start the simulations. For example, the user can run the following command to run a simulation that evolves the networks to acquire ultrasensitive response dynamics: `biojazz --config=config/ultrasensitive.cfg --tag=first_try --cluster_type="LOCAL" --cluster_size=2`. The `cluster_type` and `cluster_size` arguments override the specification contained in the configuration file, and will launch both worker nodes of the cluster on your machine. The `tag` argument is very important. In BioJazz, each design attempt is associated with a specific, user-specified tag. BioJazz will create a directory that is named with the tag in your workspace. It contains all the results and other files generated during the evolutionary simulations. This setting allows the user to attempt several simulations simultaneously without fear of accidental loss of files. The name of the design's working directory is `work_dir/tag`. The `work_dir` parameter is specified in the configuration file and has a value of template in this example. The results of the above run are contained in the directory `ultrasensitive/first_try`.

The `obj` directory contains all the genomes generated during an evolutionary simulation in a machine-readable format. The `matlab` folder contains the corresponding ANC rule-based models, reaction network models generated by ANC, and the Matlab scripts generated by Facile, these scripts can be evaluated at post-simulation stage using ANC, Facile and MATLAB accordingly. The `stat` contains the output information of each genome in each generation in `.csv` files, all the attributes are defined in the configuration file in the previous subsection. The `source_2013-06-03-14:51:58` directory is a snapshot of the source code, including the configuration and custom scoring files, used for the simulation, the time stamp at the end of its name shows the starting time of simulation. Now the user can try modifying the configuration file to use available workstations and run BioJazz. The performance of BioJazz is shown and explained in Reference *(20)*.

**Results**

To illustrate the use of BioJazz as an approach for better understanding the design principles of signaling networks, we run here evolutionary simulations using a fitness function that selects for "ultrasensitive" response dynamics *(20, 21)*. We use three pre-defined starting networks and for each of them, we run evolutionary simulations using two different parameter sets, corresponding to cellular conditions mimicking enzyme saturation or not *(21)*. We run 10 simulations for each setting with `kimura_selection` method, which results into 60 simulations in total (when using this selection method, it is possible to submit several BioJazz simulations at the same time to increase the efficiency).

*Bistability emerges from evolutionary simulations*
From the 60 simulations run, 21 evolved networks are sufficiently "ultrasensitive" (fitness score larger than 0.8) *(21)*. These networks show diverse structure patterns and parameter configurations *(21)*. Some of these networks take advantage of the so called "zero-order sensitivity" mechanism *(29)* to achieve switching-like response dynamics *(20, 21)*, while some others utilize enzyme sequestration to generate sufficient threshold and sensitivity *(21)*. Here, we report another discovered mechanism which is bistability. In the following, we will show how to discover the structural patterns for such dynamics and function.

When we study the dose response curves of all evolved ultrasensitive networks, there are 7 networks whose dose response curves show clear hysteresis near the threshold (Figure 4). From low level of output to high level of output or *vice versa*, the hysteretic transitions in these networks indicate there are two distinct stable steady states in their dynamics. When input signal is in the hysteretic area, the system has two distinct levels of output response (i.e.

they are bistable). However, which state the system stays in depends on the historical state where the system comes from *(30-34)*.

Although switching with hysteresis is not the targeted response dynamics in the evolutionary simulations, the designed fitness function selects for a wide threshold where signals response most. The hysteresis in evolved bistable networks provides sufficient threshold such that the system response mostly in the hysteretic range of input signal. Therefore, evolved bistable response dynamics is one of the possible solutions to the designed fitness function for selecting ultrasensitive response dynamics. However, the bistable dynamics is different from ultrasensitive dynamics in the sense that the latter is monostable and without hysteresis. The evolved bistable networks are rather complex in terms of combinatorial interactions between signalling proteins with multiple domains (Figure 4). It is difficult to map their structures with the underlying mechanisms where bistability emerges. In previous studies, bistable dynamics in biological systems are commonly linked to positive feedback loops which are immediately observed from schematics of gene regulatory networks *(35)*. Mathematical proofs also showed that the presence of positive feedback loops is a requirement for gene networks displaying multisationarity *(36)*. In all the evolved bistable signalling networks here, positive feedback loops that endow bistability are not directly observable, but are potentially embedded in the complex interactions (Figure 4). This observation is similar to those made from investigations on natural Mitogen Activated Protein Kinase (MAPK) signalling pathways, which show that phosphorylation and desphorylation cycles of proteins with multiple phosphorylation sites can result in multistationarity even though no obvious positive feedback loops can be found in the structure of signalling cycles *(37)*. In the evolved bistable networks here, there are no proteins with multiple phosphorylation sites and no obvious feedback interactions. Altogether these hint that evolved bistable networks embed novel interaction or dynamical motifs to enable hysteresis and bistability.

*Dissecting evolved bistable networks to decipher principles for bistability*
In order to understand such underlying motifs, we started to dissect the structure of evolved bistable networks. This is achieved by reducing the complexity of evolved networks by removing specific interactions and/or proteins one-by-one in a manual fashion, and analysing the ensuing response dynamics from the resulting simpler networks (see also Discussion). Dynamics can be analysed through temporal simulations, but also by utilising analytical tools such as the chemical reaction network toolbox (CRNToolbox) (see **Note 5**). The CRNToolbox uses chemical reaction networks theory (CRNT) *(38-41)* in order to check several qualitative properties of chemical reaction networks with mass-action kinetics. One such property is the existence of multistationarity with any parameters in the positive real domain. This parameter-free approach can help us find the minimal structure basis of multistationarity in evolved networks. In each step, we simplify the evolved networks by removing a single signalling protein or interaction in the network, then use CRNToolbox to check if the network is still bistable. The network is simplified until it becomes monostable. Then the minimal network structures are considered as candidate subnetworks enabling multistationary in evolved networks.

To implement this approach, we started from a relatively simple network (Network 15 in Figure 4) and continued simplifying the network into smaller and smaller structures that still allow bistable dynamics (Figure 5). At the first two simplification steps the toolbox cannot determine whether there are more than one steady state or not, because of limitations of the CRNT (the current algorithms limit the application of CRNT to only small networks where inequality system are linear). These systems are thus analysed through temporal simulations and bistability confirmed. All evolved networks contain allosteric regulations that did not exist from where the evolutionary simulations started (Figure 1 in *(21)*). We firstly take a route to reduce the size of network while keeping the allosteric regulations. The derived smallest bistable subnetwork is composed of one phosphorylation-dephosporylation cycle with an allosteric enzyme where the kinase has two distinct conformational states that switch between each other. Further simplifying this bistable subnetwork by removing allosteric reactions results in a subnetwork of phosphorylation-dephosphorylation cycle which is monostable. The monostable cycle is exactly the same as the well-studied zero-order sensitivity model *(29)*. This supports an hypothesis that allosteric enzymes are important for bistable dynamics in signalling networks. Inspection of all other evolved bistable networks shows that this phosphorylation-dephosphorylation cycle with allosteric enzyme exists in all evolved bistable networks. Thus, we conclude that a phosphorylation-dephosphorylation cycle featuring an allosteric enzyme is one of the simplest motifs for generating bistable dynamics in evolved signalling network and the prerequisite of bistability in this motif is the allosteric switching of kinase. The detailed mechanisms through which this motif gives rise to bistable dynamics are explained in a recent paper *(42)*.

*Design of bistable networks without allosteric regulations*
Since the evolutionary simulation with the discussed fitness function allows bistable response dynamics to occur, the fitness function can be used as an objective function to design further bistable signalling networks. This approach can allow discovery of other patterns or design principles for bistable dynamics in signalling networks. To explore this proposition, we ran another 60 simulations with the same starting conditions and fitness function as before except that no allosteric regulations were allowed to evolve in the simulations. In this setting, the bistable motifs with allosteric enzymes discovered from previously evolved networks cannot appear, forcing the evolution of other mechanisms for bistability (if possible).

These evolutionary simulations have resulted in only 3 networks that become "ultrasensitive" (fitness score larger than 0.8). However, from those 3 networks, two of them have hysteresis in their dose response curve and thus are bistable (Figure 6). This clearly shows that there are mechanisms other than allostery that are endowing the evolved networks with bistable dynamics. Using the similar deducing approach, we dissect one of these two bistable networks (Network B2) and derive a simple bistable motif (right panel in Figure 7) featuring a futile cycle with both enzymes (i.e. the kinase and phosphatase) binding each other. Since a futile cycle where the kinase and phosphatase do not interact, the well-known Goldbeter-Koshland motif *(29)*, is not bistable we conclude the crucial element for bistabilty is this binding interaction. The evolved network (B2) can also be reduced to give rise another simple network that has a kinase and phosphatase, which are then seqeuestered by (or interacting with) a single protein (left panel in Figure 7). A separate analysis has shown that, while this structure has interesting dynamical properties, and allows for both ultrasensitive and adaptive response dynamics *(21)*, it can not allow for bistability. These two simplified network motifs only differ on how the kinase and phosphatase are sequestered. Despite this minor difference, the resulting system dynamics are qualitatively different with one motif displaying multistationarity, while the other not. More analyses are needed to dissect the structural and dynamical features of these motif that give rise to this difference, however, this finding indicates that there can be very saddle features in signaling networks that can give rise to significantly different capacity for multistationarity *(43)*.

**Discussions**

Here, we have discussed the implementation of an *in silico* evolution platform for studying the evolution of structure-function relation in signalling networks. This platform combines rule-based modelling of signalling networks with an evolutionary algorithm, thereby allowing flexible and open-ended evolution of network structure. By using appropriate selection functions, we showed how this approach can be used to explore evolution of network structures that confer specific response dynamics. In particular, we found that imposing selection for switch-like response dynamics allows the discovery of networks with bistablity. Furthermore, we show that this evolutionary approach can be constrained in various ways to implement specific environmental or cellular conditions, thereby allowing the study of the impact of such constraining conditions on resulting network structures and dynamics. These results show the value of the *in silico* evolution platform for identifying networks embedding a specific system dynamics and for deciphering their key structural determinants that link structure to function, i.e. that constitute potential design principles.

Combining evolution *in silico* with rule-based modelling approach allows us to explore the evolutionary landscapes of different signaling dynamics and information processing functions. Evolved networks out of those simulations will locate the so called "solution space" for the function. On one hand, this method can be used as a design tool to optimize signaling networks specific for one or multiple functions; on the other hand, by studying the evolved networks, one can uncover different design principles for certain dynamics or functions. Moreover, the platform may allow us to explore the solutions to complex functions, for instance certain networks can install with different functions *(21)*. Therefore, it is an intriguing question that how the signalling networks evolve when one increases the complexity of functions that the signalling networks are selected for, which mimics increasing the complexity of environments where the cell sits in. Such investigations may help us understand the design principles of complex information processing function as well as the origination of complexity in signalling networks. However, in most cases those evolved networks are still relatively complex due to the unbounded complexity in the evolutionary simulations. There are three feasible approaches to uncover the underlying structure patterns that are comprehensible and intuitive. The first approach is merely by observing and comparing the structures of evolved networks to get hypothetical structure patterns then using synthesized small networks to validate the hypothesis and derived predictions *(21)*. The second approach is manually reducing the size and simplifying the structure of evolved networks while using certain tools to determine the maintenance of functions, as shown in this study (see **Results**). A third approach would be extending the current evolutionary rule-based models to continue evolve the resulted networks with harsh constraints on the complexity of networks, this may result into many different structure patterns for maintaining certain functions. This would provide a more systematic way to discover design principles for system dynamics, which can be implemented in the future work.

Evolvable rule-based models are not just applicable to studying signaling networks, and the presented approach can be applied to other cellular networks. The combinatorial complexity in signaling networks also exists in gene regulatory networks *(44)*, chromatin modifications *(45)* and transcriptional controls *(46)*. Therefore, the approaches discussed here can also be used to explore design principles of gene regulatory networks. It is particularly intriguing that signaling networks have a tight interface with gene regulatory networks. By extending current approach for signaling networks to an integrated regulatory system, it is possible to understand the fundamental design principles of cellular decision makings *(47)* in which cells response properly even when exposed to many combinatorial environmental signals.

**Notes**

1. BioJazz, ANC and Facile requires several Perl modules to run properly, which might not be readily installed in the machine. CPAN is an internet database of Perl modules. The user will need system administrator privileges to install these modules or see for instructions on how to install them in your home directory (http://twiki.org/cgi-bin/view/TWiki/HowToInstallCpanModules). If with administrator privileges, the user typically need to run the following commands:

   ```
   sudo cpan -i Class::Std
   sudo cpan -i Class::Std::Storable
   sudo cpan -i String::CRC32
   sudo cpan -i Expect
   sudo cpan -i Carp
   sudo cpan -i WeakRef
   sudo cpan -i IPC::Shareable
   sudo cpan -i Linux::Pid
   sudo cpan -i Text::CSV
   sudo cpan -i GraphViz
   ```

2. With all components installed, you can tell BioJazz where to get them by setting the `ANC_HOME` and `FACILE_HOME` environment variables to point to the appropriate directories. It is recommended to add the following lines to your "`~/.bashrc`" or "`~/.bash_profile`" file file:

   ```
   export ANC_HOME=~/workspace/anc
   export FACILE_HOME=~/workspace/facile
   alias  anc='$ANC_HOME/anc.pl'
   alias  facile='$FACILE_HOME/facile.pl'
   export BIOJAZZ_HOME=~/workspace/biojazz
   alias  biojazz='$BIOJAZZ_HOME/biojazz.pl'
   ```

   BioJazz requires Matlab to be installed on all nodes used for computation, and assumes matlab can be started with the command "`matlab`". Here is an example of configuration in "`~/.bashrc`" file or "`~/.bash_profile`" file on Linux (Ubuntu):

   ```
   export MATLAB_HOME=/usr/local/MATLAB/R2015b/bin
   alias  matlab='$MATLAB_HOME/matlab'
   export PATH=$MATLAB_HOME:$PATH
   DYLD_LIBRARY_PATH=/usr/local/MATLAB/R2015b/bin/maci64:/usr/local/MATLAB/R2015b/sys/os/maci64:/usr/local/MATLAB/R2015b/runtime/maci64:$DYLD_LIBRARY_PATH
   export DYLD_LIBRARY_PATH
   ```

3. If the user wants to study alternative cellular information processing functions, like oscillatory dynamics, pulsatile responses or even combinatorial inputs and outputs, at the current stage of development the user should implement such fitness functions on their own. However, the design of the BioJazz provide such flexibility to program desired fitness functions. Relevant information and interfaces can be found in the source code.

4. The radius parameter should be reasonable in the sense that it will maintain adequate promiscuity of protein-protein interactions. If the radius is too small, protein-protein interactions are more restricted thus the resulting signaling networks are sparser, whereas if the radius is too large, protein-protein interactions are more promiscuous and the signaling networks are more complex. One can also try to evolve the radius or modify the radius value to study the roles of protein promiscuity in evolution of signaling networks.

5. For determining the existence of multistationarity of given signalling networks, I utilised the Chemical Reaction Network Toolbox (CRNToolbox, https://crnt.osu.edu/CRNTWin). Given a chemical reaction network described with mass action kinetics, CRNToolbox can determine whether multiple equilibria exist with any positive kinetic parameters. I analysed the existence of multistationarity in several different signalling networks given the chemical reactions in the networks. An example of CRNToolbox report can be found in (Supplementary file). The detail usage of CRNToolbox is described in its manual.

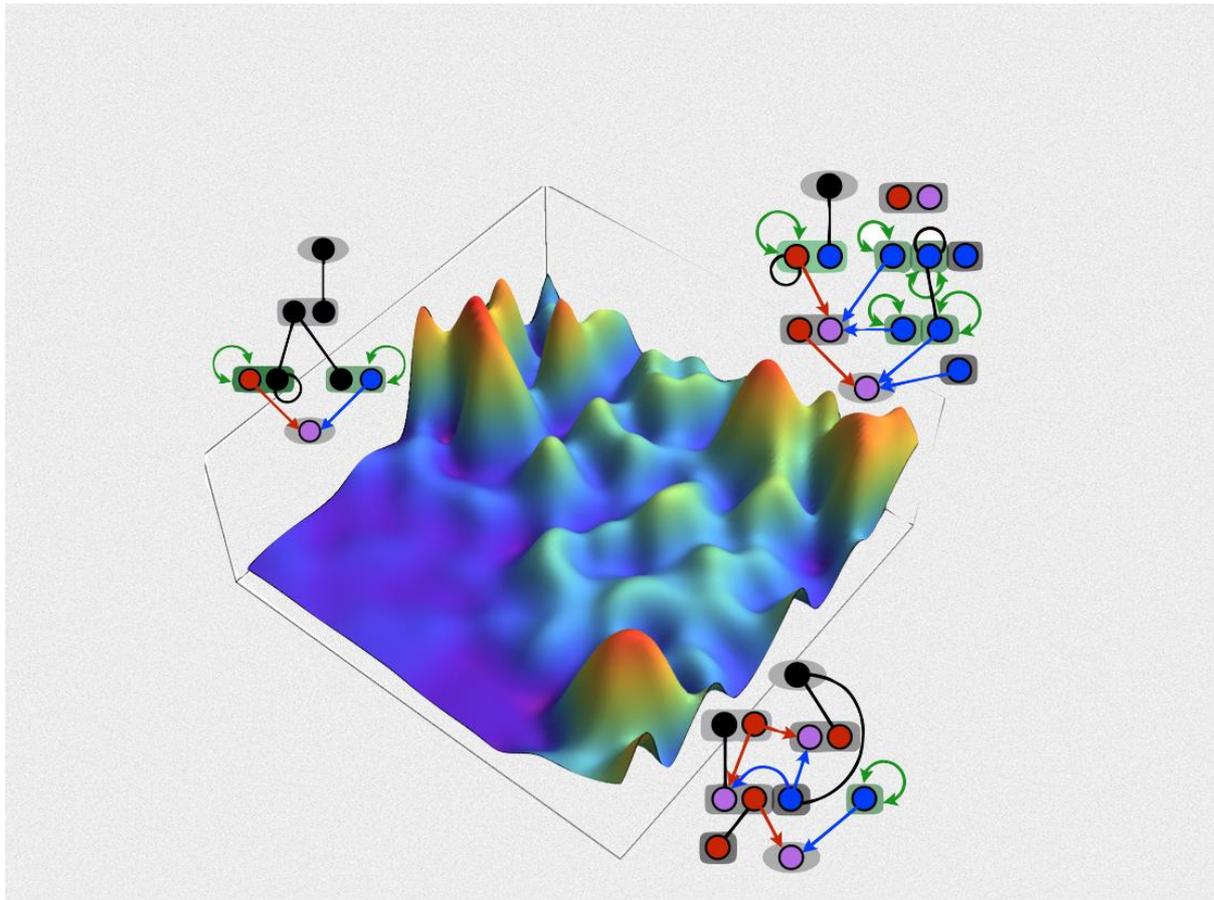

Figure 1. A schematic evolutionary landscape of signaling networks. The altitude of peaks showing the different solutions with high fitness in such landscapes.

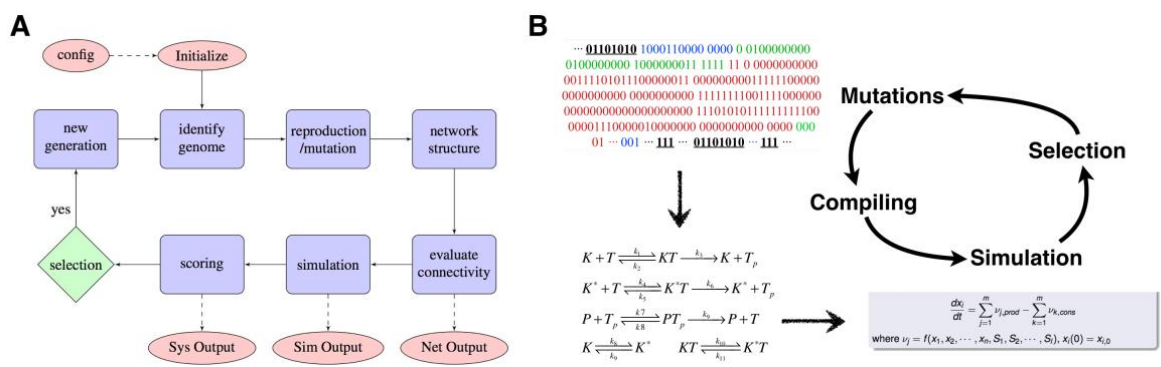

Figure 2. A schematic of evolutionary algorithm implemented in BioJazz. (A) The simulation steps of evolutionary algorithm. The whole algorithms include an initiation step and iteration cycles until the termination conditions are met. (B) For each evolutionary iteration, the format of the rule-based model. At each iteration, the binary string are mutated then complied into rules and consequently differential equations, then all the mutants will be scored by fitness function and finally the next generation will result from selection on the currently generation.

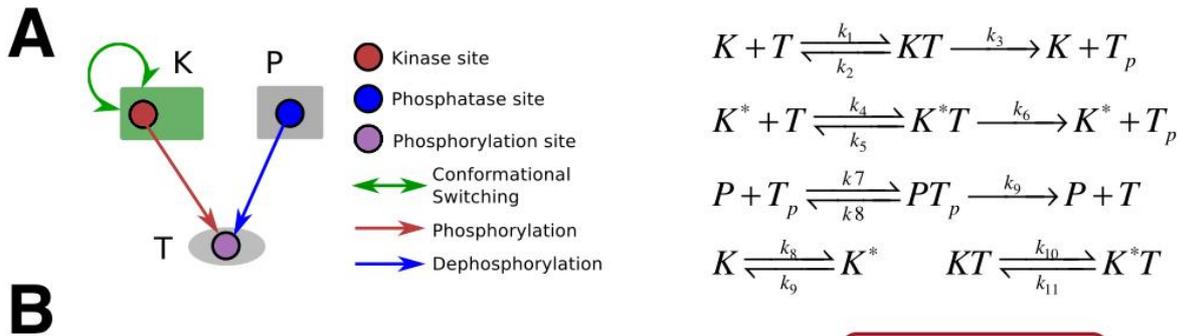

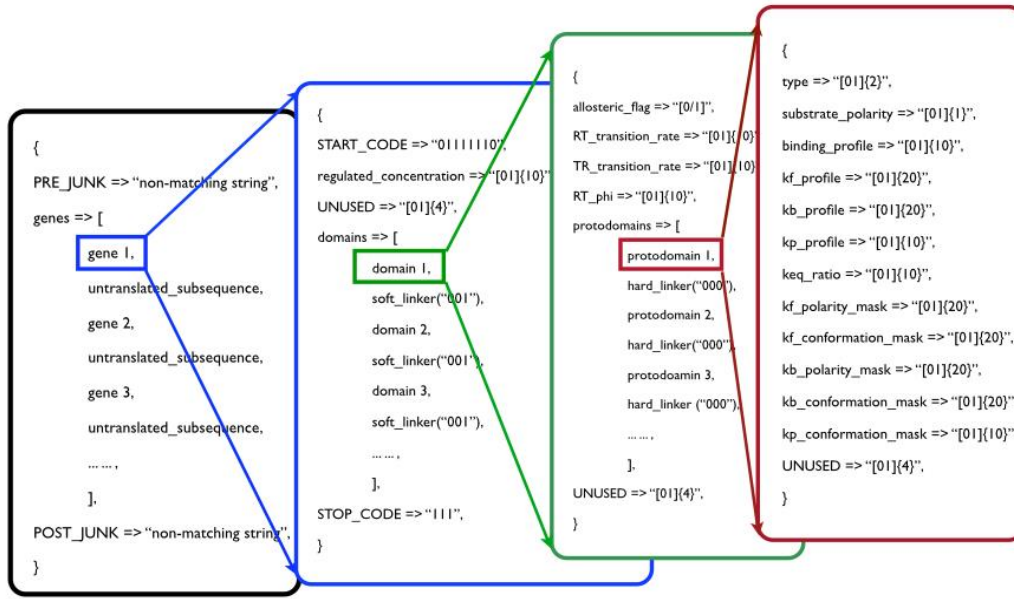

Figure 3. The schematic structure of networks and the actual structure implemented in the program. (A) The schematic structure of a signaling network with its according chemical reactions. (B) The actual structure implemented in BioJazz. The network is encoded in a binary strings with hierarchical structures. Each rectangular contains a layer of structure, from "genome" to "proteins" then to "domains" and "reactive sites" *(20)*.

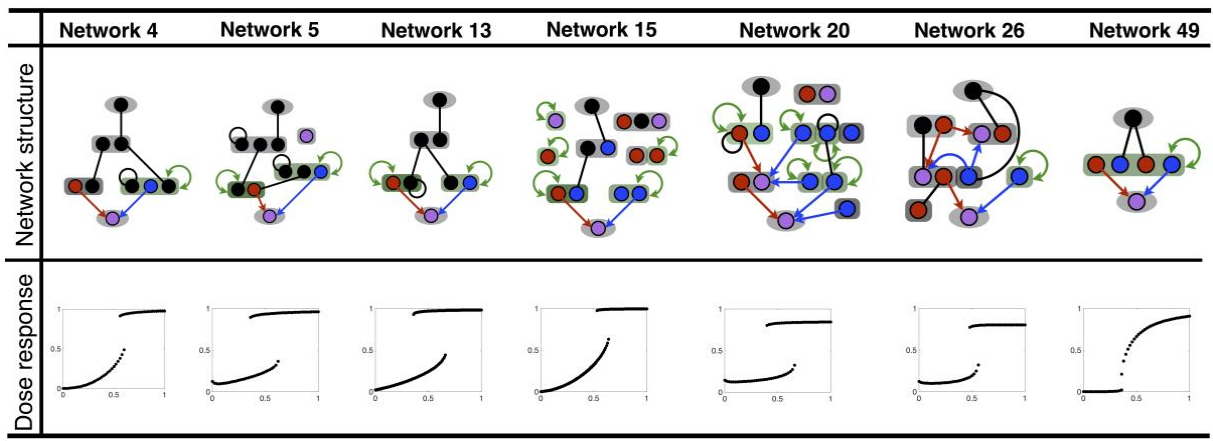

Figure 4. The emerged networks with hysteresis in their switch-like response dynamics. The picture shows the structure of the all seven networks and their dose-response curves.

* This is a pre-peer-review, pre-copyedit version of an article that will appear as a chapter in Modelling Molecular Site Dynamics. Editors: Hlavacek, William S. (Ed.). Publisher: Springer

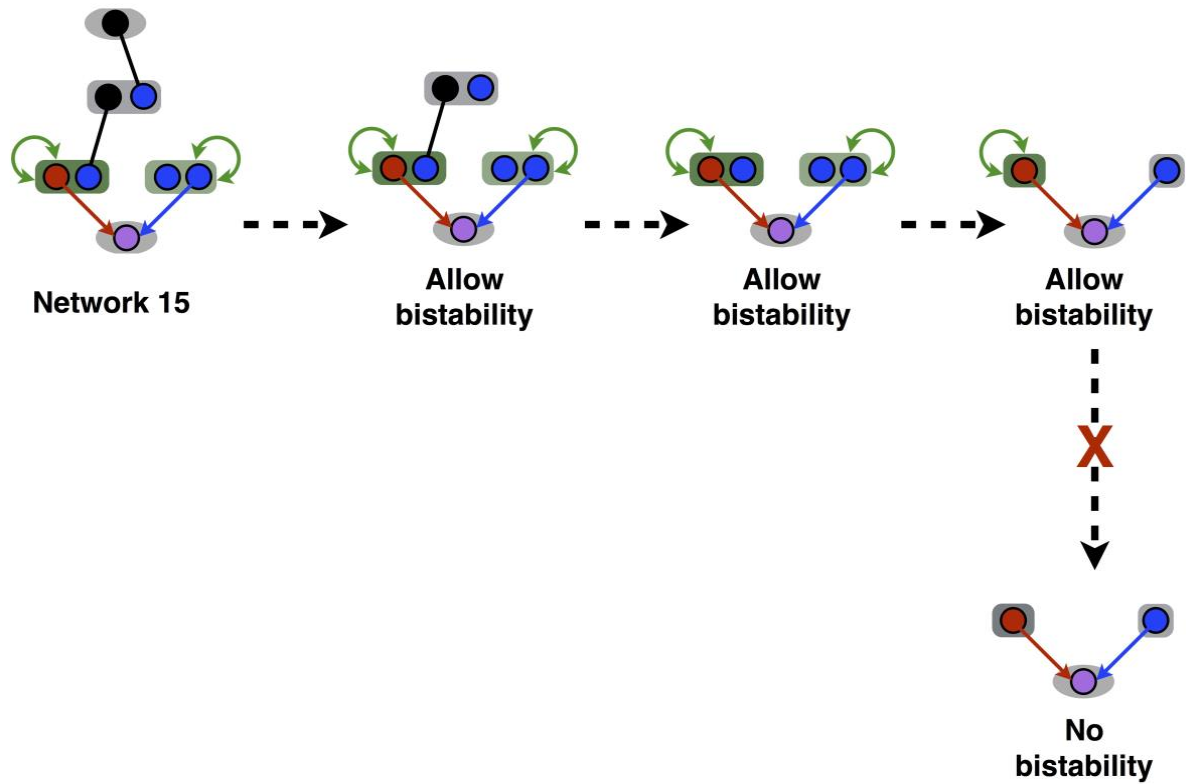

Figure 5. Simplifying the evolved Network 15 from Figure 4 to the simplest networks that still maintain bistability.

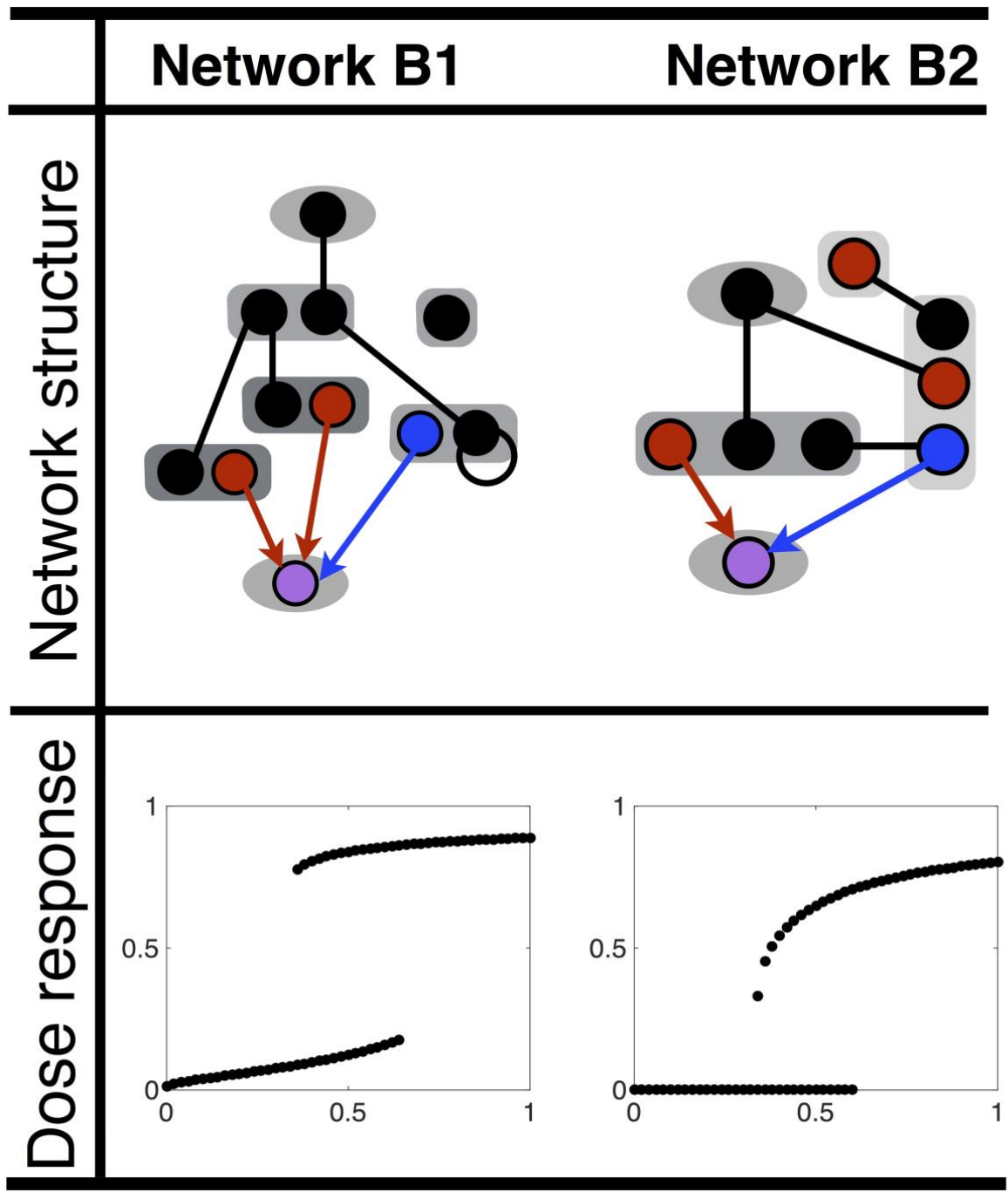

Figure 6. Evolved networks with hysteresis in their switch-like response from evolutionary simulations where allosteric regulations are not allowed to emerge.

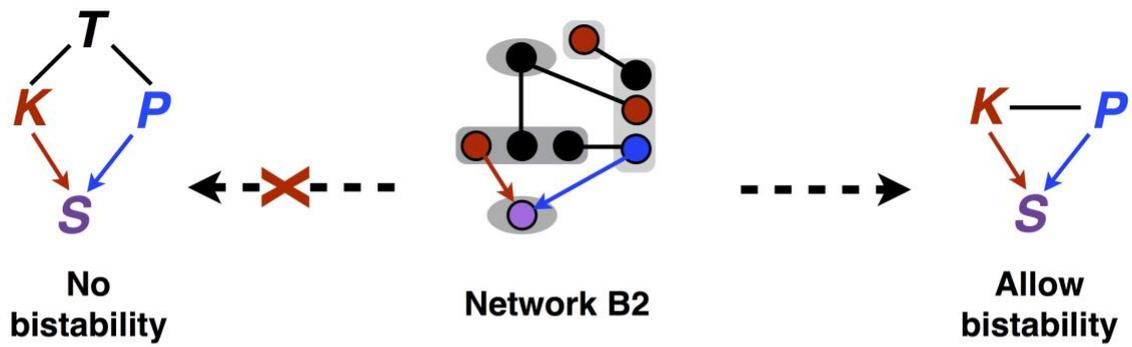

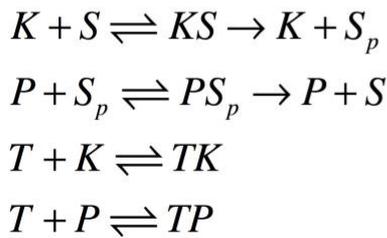

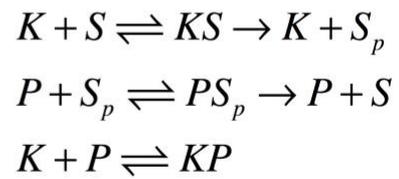

$$K + S \rightleftharpoons KS \rightarrow K + S_p$$
$$P + S_p \rightleftharpoons PS_p \rightarrow P + S$$
$$T + K \rightleftharpoons TK$$
$$T + P \rightleftharpoons TP$$

$$K + S \rightleftharpoons KS \rightarrow K + S_p$$
$$P + S_p \rightleftharpoons PS_p \rightarrow P + S$$
$$K + P \rightleftharpoons KP$$

Figure 7. Simplification of Network B2 from Figure 6 to a simplest network motif that maintains the capacity of bistability (right) and to another motif that has no capacity of bistability (left). The two motifs only differ at whether the sequestration of kinase and phosphatase by each other (right) or by an external sequestrating protein (left).